\documentclass[conference]{IEEEtran}

\usepackage{amsfonts,amsbsy,bm,amsmath}
\usepackage[nolist]{acronym}
\usepackage{mathrsfs} 
\usepackage{graphicx,cite,amssymb,amsmath,mathtools,amsthm}
\usepackage{tikz}
\usetikzlibrary{fadings}
\usetikzlibrary{shadows.blur}
\usetikzlibrary{shapes,arrows}
\usetikzlibrary{calc}
\usetikzlibrary{decorations.pathreplacing}
\usepackage{ifthen}
\usetikzlibrary{positioning}
\usepackage{caption}
\usepackage{subcaption}

\usepackage{pgfplots,relsize}
\usepackage{color}

\usepackage[font=small]{caption}

\usepackage{cleveref}
\usepackage{bbm}
\allowdisplaybreaks

\newcommand{\sss}[1]{\scalebox{1.1}{$\scriptscriptstyle \mathsf{#1}$}}
\newcommand{\List}{\mathcal{L}}

\newcommand{\code}{\mathcal{C}}

\newcommand{\vecx}{\boldsymbol{x}}
\newcommand{\vech}{\boldsymbol{h}}

\newcommand{\vecn}{\boldsymbol{n}}

\newcommand{\vecy}{\boldsymbol{y}}

\newcommand{\I}{\boldsymbol{I}}

\newcommand{\vecxd}{\bm{x}^{\sss{d}}}
\newcommand{\vecxp}{\bm{x}^{\sss{p}}}
\newcommand{\vecyd}{\bm{y}^{\sss{d}}}
\newcommand{\vecyp}{\bm{y}^{\sss{p}}}

\newcommand{\randvecxd}{\bm{X}^{\sss{d}}}
\newcommand{\randvecyd}{\bm{Y}^{\sss{d}}}

\newcommand{\argmax}[1]{\underset{#1}{\mathrm{arg \, max}}\,}
\newcommand{\argmin}[1]{\underset{#1}{\mathrm{arg \, min}}\,}

\newcommand{\nc}{n_c}

\newcommand{\randvech}{\bm{H}}

\newcommand{\randvecx}{\bm{X}}
\newcommand{\randvecy}{\bm{Y}}

\newcommand{\lro}[1]{\left({#1}\right)}
\newcommand{\lrho}[1]{\left[ {#1} \right]}

\newcommand{\norm}[1]{\lVert#1\rVert}
\newcommand{\inner}[1]{\langle#1\rangle}

\newcommand{\prob}[1]{\ensuremath{\mathbb{P}\lrho{#1}}}

\newcommand{\vecnorm}[1]{\lVert#1\rVert}

\begin{acronym}
	\acro{AWGN}{additive white Gaussian noise}
	\acro{B-AWGN}{binary input additive white Gaussian noise}
	\acro{B-DMC}{binary-input discrete memoryless channel}
	\acro{B-DMSC}{binary-input discrete memoryless symmetric channel}
	\acro{B-MSC}{binary-input memoryless symmetric channel}
	\acro{BCJR}{Bahl, Cocke, Jelinek, and Raviv}
	\acro{BEC}{binary erasure channel}
	\acro{BER}{bit error rate}
	\acro{BLEP}{block error probability}
	\acro{BP}{belief propagation}
	\acro{BPSK}{binary phase shift keying}
	\acro{BSC}{binary symmetric channel}
	\acro{CER}{codeword error rate}
	\acro{CN}{check node}
	\acro{CRC}{cyclic redundancy check}
	\acro{CSI}{channel state information}
	\acro{DE}{density evolution}
	\acro{EM}{expectation maximization}
	\acro{GLRT}{generalized likelihood ratio test}
	\acro{IO-WE}{input-output weight enumerator}
	\acro{i.i.d.}{identical and independently distributed}
	\acro{IR-WE}{input-redundancy weight enumerator}
	\acro{JIO-WE}{joint input-output weight enumerator}
	\acro{JIR-WE}{joint input-redundancy weight enumerator}
	\acro{JWE}{joint weight enumerator}
	\acro{LDPC}{low-density parity-check}
	\acro{LLR}{log-likelihood ratio}
	\acro{MAP}{maximum a posteriori}
	\acro{ML}{maximum likelihood}
	\acro{OFDM}{orthogonal frequency-division multiplexing}
	\acro{OSD}{ordered-statistics decoder}
	\acro{PSK}{phase shift keying}
	\acro{QPSK}{quadrature phase shift keying}
	\acro{PC}{Product code}
	\acro{PAT}{pilot-assisted transmission}
	\acro{pdf}{probability density function}
	\acro{RHS}{right-hand side}
	\acro{RBFC}{Rayleigh block fading channel}
	\acro{RM}{Reed-Muller}
	\acro{RV}{random variable}
	\acro{SISO}{single-input single-output}
	\acro{SPC}{single parity-check}
	\acro{SC}{successive cancellation}
	\acro{SCC}{super component codes}
	\acro{SCL}{successive cancellation list}
	\acro{SNR}{signal-to-noise ratio}
	\acro{UB}{union bound}
	\acro{VN}{variable node}
	\acro{WE}{weight enumerator}
	\acro{BLER}{block error rate}
	\acro{RCUs}{random-coding union bound with parameter $s$}
\end{acronym}
\IEEEoverridecommandlockouts
\begin{document}

	\title{Low-Complexity Joint Channel Estimation and List Decoding of Short Codes}

	\author{\IEEEauthorblockN{Mustafa Cemil Co\c{s}kun\IEEEauthorrefmark{1}\IEEEauthorrefmark{2},
			Gianluigi Liva\IEEEauthorrefmark{1},
			Johan \"{O}stman\IEEEauthorrefmark{3} and 
			Giuseppe Durisi\IEEEauthorrefmark{3}}
		\IEEEauthorblockA{\IEEEauthorrefmark{1} Institute of Communications and Navigation, German Aerospace Center, We{\ss}ling, Germany.
		}
		\IEEEauthorblockA{\IEEEauthorrefmark{2} Institute for Communications Engineering, Technical University of Munich, Munich, Germany
		}
		\IEEEauthorblockA{\IEEEauthorrefmark{3}Department of Electrical Engineering, Chalmers University of Technology, Gothenburg, Sweden
		}
		\thanks{This work was supported in part by the research grant "Efficient Coding and Modulation for Satellite Links with Severe Delay Constraints" funded by Munich Aerospace e.V., and by the Swedish Research Council, under grants $2014-6066$ and $2016-03293$.}
	}

	\maketitle
	
	\begin{abstract}
		A pilot-assisted transmission (PAT) scheme is proposed for short blocklengths, where the pilots are used only to derive an initial channel estimate for the list construction step. The final decision of the message is obtained by applying a non-coherent decoding metric to the codewords composing the list. This allows one to use very few pilots, thus reducing the channel estimation overhead. The method is applied to an ordered statistics decoder for communication over a Rayleigh  block-fading channel. Gains of up to $1.2$ dB as compared to traditional PAT schemes are demonstrated for short codes with QPSK signaling. The approach can be generalized to other list decoders, e.g., to list decoding of polar codes.
	\end{abstract}
	
	\section{Introduction}\label{sec:intro}

The interest in designing wireless communication systems with short information blocks, up to a few tens of bits, has been increasing recently due to the rise of  applications characterized by strict latency constraints \cite{Durisi16:Short}. As a consequence, the fundamental limits of communications for finite-length messages have received renewed attention (see, e.g., \cite{Dolinar98:BOUNDS,Fossorier04:BOUNDS,SasonShamai06:BOUNDS,polyanskiy10-05a}). Code designs \cite{Poulliat08:BinImag,Liva13:ShortTC,Dolecek14:NBLDPCTIT} and sophisticated decoding algorithms \cite{FosLin95,Tal15:ListPolar} targeting near-optimal performance in the moderate- and short-length regimes have been proposed. Using such methods, it is possible to operate close to the finite length bounds (see, e.g., \cite{Coskun18:Survey} for a comparison of short code constructions and finite length bounds). While most of the attention has been focused on communication over \ac{AWGN} channels, it is also interesting to communicate with short packets over a fading channel with no \ac{CSI} available at the transmitter and receiver. In fact, classic \ac{PAT} methods \cite{Tong04:PAT} turn out to be highly sub-optimal when short blocks are used \cite{Liva17:Mismatched}.
The rates achievable over fading channels when the \ac{CSI} is not available a priori has been investigated in \cite{Yang14:QSFC,Durisi16:MRFC,östman17-02a} for a fixed blocklength and error probability. Bounds on the error probabilities are provided in \cite{johan:unpublished} not only for non-coherent transmission but also for \ac{PAT} strategies.
 
We extend the work of \cite{johan:unpublished} by introducing a \ac{PAT} scheme with very few pilot symbols. The pilot symbols are used to obtain a (potentially rough) channel estimate, which is then employed by a list decoder to explore the neighborhood of the channel observation, i.e., to construct a list of candidate codewords that achieve a large likelihood given the available channel estimate. The final decision is then performed by selecting the codeword in the list according to a non-coherent decoding metric. The role of the pilot symbols is thus to enable the construction of a good list---a task that is less challenging than deriving directly a decision on the transmitted codeword. This enables one to allocate very few pilots, hence reducing the pilot overhead. The principle can be applied to list decoders in general and to various slow fading channels. As an example, we apply the method to the \ac{OSD} of \cite{FosLin95} over a \ac{SISO} Rayleigh block-fading channel.
 
The paper is organized as follows. In Section \ref{sec:prelim}, we present the system model and various decoding criteria, and we discuss the complexity of non-coherent decoding metrics. Motivated by the complexity argument, we review classic \ac{PAT} approaches in Section \ref{sec:pragmatic}. A list decoding method is presented in Section~\ref{sec:opt_in_list}. Finite-length performance bounds are provided in Section \ref{sec:bounds}, followed by numerical results and conclusions in Sections \ref{sec:results} and \ref{sec:conc}, respectively.
	\section{Preliminaries}\label{sec:prelim}

We use capital letters, e.g., $X$, for \acp{RV} and their lower case counterparts, e.g., $x$, for their realizations. We denote the random vectors via capital bold letters, e.g., $\randvecx = [X_1, X_2,\dots,X_n]$, and their vector realizations via the lower case counterparts, e.g., $\vecx = [x_1,x_2,\dots,x_n]$. As an exception, $\I_a$ refers to the $a\times a$ identity matrix. The probability mass function of the discrete \ac{RV} $X$ is denoted as $P_X$, whereas the probability density function of the continuous \ac{RV} $X$ is denoted as $p_X$. We use $\norm{\cdot}$ for the $l^2$-norm, $\inner{\cdot,\cdot}$ for the inner product of two vectors, $\ln(\cdot)$ for the natural logarithm, and $\mathbb{E}[\cdot]$ for the expectation. We write $\mathcal{CN}(\mu,\sigma^2)$ to denote a complex Gaussian \ac{RV} with mean $\mu$ and variance $\sigma^2$.

\subsection{System Model}
We consider a \ac{SISO} Rayleigh block-fading channel, i.e., the random fading coefficient is constant for $n_c$ channel uses and changes independently across $\ell$ coherence blocks, which are also called diversity branches. Therefore, the packet size is $n = \ell n_c$. Such a setup is relevant for \ac{OFDM} systems, e.g., LTE and 5G (see \cite{östman17-02a}).
The input-output relationship of the channel for the $i$th coherence block is
\begin{equation}\label{eq:model}
	\vecy_i = h_i\vecx_i+\vecn_i, \quad i = 1,\dots,\ell
\end{equation}
where $\vecx_i\in\mathcal{X}^{n_c}$ and $\vecy_i\in\mathbb{C}^{n_c}$ denote the transmitted and received vectors, $h_i$ is the realization of the channel coefficient, which is distributed as $H_i\sim\mathcal{CN}(0,1)$ and $\vecn_i$ is the corresponding \ac{AWGN} term, which is distributed as $\boldsymbol{N}_i\sim\mathcal{CN}(\boldsymbol{0},\sigma^2\boldsymbol{I}_{n_c})$.  The mutually independent \acp{RV} $H_i$ and $\boldsymbol{N}_i$ are assumed to be independent over $i$.
We will focus on \ac{QPSK} signalling where energy per symbol is normalized to $1$.

\subsection{Decoding with Perfect \ac{CSI}}\label{subsec:perfect}
If the channel coefficients are known to the receiver, the (coherent) \ac{ML} decoding rule is
\begin{align}\label{eq:coh_ML}
	\hat{\vecx} &= \argmax{\vecx\in\code}p_{\randvecy|\randvecx,\randvech}(\vecy|\vecx,\vech) \\
	&= \argmin{\vecx\in\code}\sum_{i=1}^{\ell}||\vecy_i-h_i\vecx_i||^2 
\end{align}
where $\code$ is the set of transmitted signal vectors induced by the chosen channel code and modulation. When $\lVert\vecx_i\rVert$ is constant across codewords and blocks, we have
\begin{equation}
	\hat{\vecx} = \argmax{\vecx\in\code}\sum_{i=1}^{\ell} \Re\{\langle\vecy_i,h_i\vecx_i\rangle\}
\end{equation}
which is the case, for instance, if the modulation is \ac{QPSK}.

\subsection{Decoding without \ac{CSI}---Pilot-Assisted Channel Estimation}\label{subsec:PAT}
The idealized setting described in Section \ref{subsec:perfect} is often approximated by including pilot symbols in the transmitted sequence, which are used to obtain an estimate of the channel coefficients. This estimate $\hat{\vech}$ is treated as ideal by a mismatched decoder, yielding
\begin{align}
	\hat{\vecx} &= \argmax{\vecx\in\code}p_{\randvecy|\randvecx,\randvech}(\vecy|\vecx,\hat{\vech}) \\
	&= \argmin{\vecx\in\code}\sum_{i=1}^{\ell}||\vecy_i-\hat{h}_i\vecx_i||^2. \label{eq:PAT}
\end{align}
For \ac{QPSK}, this reduces to
	\begin{equation}\label{eq:PAT_QPSK}
	\hat{\vecx} = \argmax{\vecx\in\code}\sum_{i=1}^{\ell} \Re\{\langle\vecy_i,\hat{h}_i\vecx_i\rangle\}.
\end{equation}

\subsection{Decoding without \ac{CSI}---Blind Approach}
Assume next that the decoder does not have access to the channel coefficients, and that no pilots are embedded in the transmitted sequence. In this case, we distinguish between two possibilities: $i)$ The decoder does not possess information  on the distribution of the channel coefficients and $ii)$ the decoder knows the channel coefficients' distribution. 
In case $i)$, the problem can be tackled, for instance, by designing a \ac{GLRT} as in \cite{Warrier02} yielding
\begin{align}
	\hat{\vecx} &= \argmax{\vecx\in\code}\sup_{\vech}p_{\randvecy|\randvecx,\randvech}(\vecy|\vecx,\vech) \\
	&= \argmin{\vecx\in\code}\sum_{i=1}^{\ell}\inf_{h_i}||\vecy_i-h_i\vecx_i||^2 \\
	&=\argmax{\vecx\in\code}\sum_{i=1}^{\ell}\frac{|\langle\vecy_i,\vecx_i\rangle|^2}{\lVert\vecx_i\rVert^2}.\label{eq:GLRT}
\end{align}
The last step follows because the \ac{ML} channel estimate is $\hat{h}_{i} = \inner{\vecy_i,\vecx_i}/\norm{\vecx_i}^2$. For \ac{QPSK}, \eqref{eq:GLRT} reduces to
\begin{equation}\label{eq:GLRT_QPSK}
\hat{\vecx} = \argmax{\vecx\in\code}\sum_{i=1}^{\ell}|\langle\vecy_i,\vecx_i\rangle|^2.
\end{equation}

In case $ii)$, the non-coherent \ac{ML} estimate is
\begin{align}
	\hat{\vecx} &= \argmax{\vecx\in\code} \prod_{i=1}^{\ell} \mathbb{E}[p_{\randvecy|\randvecx,H}(\vecy_i|\vecx_i,H)] \label{eq:nc_ML_def}\\
	&= \argmax{\vecx\in\code} \sum_{i=1}^{\ell} \frac{|\langle\vecy_i,\vecx_i\rangle|^2}{\sigma^2(\sigma^2+\norm{\vecx_i}^2)}-\ln\left(1+\frac{\norm{\vecx_i}^2}{\sigma^2}\right)
	\label{eq:nc_ML}
\end{align}
where \eqref{eq:nc_ML} follows because the conditional received vector $\vecy_i$ per coherence block given the transmitted sequence $\vecx_i$ is complex Gaussian with mean $\mathbb{E}[\vecy_i|\vecx_i] = \boldsymbol{0}$ and covariance $\mathbb{E}[\vecy_i^H\vecy_i|\vecx_i] = \sigma^2\I_{n_c}+\vecx_i^H\vecx_i$. For QPSK, we get
\begin{equation}
	\hat{\vecx} = \argmax{\vecx\in\code} \sum_{i=1}^{\ell} |\langle\vecy_i,\vecx_i\rangle|^2.
	\label{eq:nc_ML2}
\end{equation}
Note that \eqref{eq:GLRT_QPSK} coincides with \eqref{eq:nc_ML2} under the assumption that the signals in $\code$ have the same energy over each coherence block, e.g., for \ac{QPSK} signaling.

\subsection{On the Complexity of Non-Coherent Decoding}\label{subsec:complexity}

By inspecting \eqref{eq:GLRT_QPSK} and \eqref{eq:nc_ML2}, we see that the decoding metric does not admit a trivial factorization, hindering the use of efficient maximum metric decoders (such as Viterbi decoding over the code trellis) and of any decoding algorithm that relies on the factorization of the channel likelihood (such as belief propagation decoding for turbo/low-density parity-check codes or successive cancellation decoding of polar codes).  
A pragmatic solution to this problem is to embed a small number of pilots in the transmitted sequence, which are then used to bootstrap iterative decoding and channel estimation algorithms \cite{henkbook,herzet2007code,KhaBout:EM-APP06}

In the following, we first discuss how the iterative decoding and channel estimation approach can be applied to \ac{OSD} (Section \ref{sec:pragmatic}). Then, we show that list decoders in general, and \ac{OSD} in particular, allow for an alternative approach to non-coherent decoding (Section \ref{sec:opt_in_list}) which yields simultaneously a gain in error correction capability and a reduction in decoding complexity.

The general framework for the algorithms presented in the following sections relies on the \ac{PAT} approach of Section~\ref{subsec:PAT}. More specifically, we embed $n_p$ pilot symbols into each coherence block. For the $i$th coherence block, the vector of pilot symbols is denoted by $\vecxp_i$. The pilots are followed by $n_c-n_p$ coded symbols, denoted by $\vecxd_i$. The corresponding channel outputs are denoted by $\vecyp_i$ and $\vecyd_i$, respectively. The rate in terms of bits per channel use (bpcu) is 
	\begin{equation}
		R=\frac{k}{\ell n_c}
	\end{equation}
	where $k$ is the number of information bits encoded by $\code$. The rate of the code $\code$ is instead denoted by 
	\begin{equation}
	R_0=\frac{k}{\ell (n_c-n_p)}.
	\end{equation}
	As a result, for a fixed rate $R$ and a fixed blocklength $\ell n_c$, a large number of pilots comes at the cost of an increase in the code rate $R_0$, and thus a reduction of the error correction capability. This yields a trade-off between resources allocated to channel estimation and error correction (see \cite{Liva17:Mismatched}). 
	\section{Classic Approaches}\label{sec:pragmatic}

We illustrate two ways of using OSD, which will be taken as references for the novel algorithm presented in Section \ref{sec:opt_in_list}. The first approach is a plain application of the \ac{PAT} scheme sketched in Section \ref{subsec:PAT}. The second approach iterates pilot-aided channel estimation and \ac{OSD} by means of the \ac{EM} algorithm. Upon observing the channel output, both approaches use the pilot symbols in each coherence block to perform an \ac{ML} estimation of the corresponding channel coefficient, i.e., we have
\begin{equation}\label{eq:channel_estimate}
	\hat{h}_i = \frac{\inner{\vecy_i^{\sss{p}},\vecx_i^{\sss{p}}}}{\norm{\vecx_i^{\sss{p}}}}, \quad i = 1,\dots,\ell.
\end{equation}

\subsection{Pragmatic Pilot-Assisted Ordered-Statistics Decoder}\label{sec:PATOSD}

The channel estimates \eqref{eq:channel_estimate} are treated as perfect and the bit-wise \acp{LLR} based on the mismatched likelihoods $p_{\randvecy|\randvecx,H}(\vecyd_i|\vecxd_i,\hat{h}_i)$, with $i=1,\ldots,\ell$, are fed to the \ac{OSD}.
After constructing the list $\List$, one applies the metric in \eqref{eq:PAT_QPSK} to the codewords in the list, yielding
\begin{align}\label{eq:mismatched_ML}
	\hat{\vecx}^{\sss{d}} &= \argmax{\vecxd\in\List}\sum_{i=1}^{\ell}\Re\{\langle\vecyd_i,\hat{h}_i\vecxd_i\rangle\}.
\end{align}

\subsection{Iterative Channel Estimation and Ordered-Statistics Decoding via Expectation-Maximization}

We reduce the number of pilots (and hence allow for the use of a lower-rate code) by iterating channel estimation and channel decoding \cite{henkbook,herzet2007code}. In the following, we describe how the \ac{EM} algorithm \cite{Dempster77maximumlikelihood} can be used for this purpose, in combination with \ac{OSD}. The algorithm works as follows:
\medskip

\begin{itemize}
	\item[1.] Initialize $\hat{h}_i^{(0)}$ as in \eqref{eq:channel_estimate} for $i = 1,\dots,\ell$, and construct the list $\mathcal{L}^{(0)}$ using the channel estimates.
	\item[2.] At iteration $j$, we construct the list $\mathcal{L}^{(j)}$ using the updated channel estimates $\hat{\vech}^{(j)}$. Then, we have
	\begin{itemize}
		\item[a.] Expectation step:
			\begin{align}
				Q(h_i,\hat{\vech}^{(j-1)}) &= \sum_{\vecxd\in\mathcal{L}^{(j-1)}}-P_{\randvecxd|\randvecyd,\randvech}(\vecxd|\vecyd,\hat{\vech}^{(j-1)}) \nonumber\\
				&\quad\qquad\times\norm{\vecyd_i-h_i\vecxd_i}^2
		\end{align}
		where we approximate $P_{\randvecxd|\randvecyd,\randvech}(\vecxd|\vecyd,\hat{\vech}^{(j-1)})$ as
			\begin{equation}
				 \frac{p_{\randvecyd|\randvecxd,\randvech}(\vecyd|\vecxd,\hat{\vech}^{(j-1)})}{\sum_{\tilde{\vecx}^{\sss{d}}\in\mathcal{L}^{(j-1)}} p_{\randvecyd|\randvecxd,\randvech}(\vecyd|\tilde{\vecx}^{\sss{d}},\hat{\vech}^{(j-1)})}.
			\end{equation}
		\item[b.]  Maximization step:
		\begin{equation}
			\hat{h}_i^{(j)} = \argmax{h_i} Q(h_i,\hat{\vech}^{(j-1)}).
		\end{equation}
	\end{itemize}
\end{itemize}
After performing Step 2 for a predetermined number $m$ of iterations, the final decision is obtained as in \eqref{eq:mismatched_ML} by replacing $\hat{\vech}$ and $\mathcal{L}$ by $\hat{\vech}^{(m)}$ and $\mathcal{L}^{(m)}$, respectively.
	\section{Ordered-Statistics Decoding with in-List \ac{GLRT}}\label{sec:opt_in_list}

We use the channel estimate to form the list $\mathcal{L}$ of codewords via the \ac{OSD} procedure as in Section \ref{sec:PATOSD}. Then, each codeword in the list is modified by re-inserting the pilot field, which yields a modified list $\List'$. The final codeword is picked among $\List'$ according to the \ac{GLRT} rule given by \eqref{eq:GLRT_QPSK}, i.e., we choose
	\begin{align}
	\hat{\vecx} &= \argmax{\vecx\in\List'}\sum_{i=1}^{\ell}|\langle\vecy_i,\vecx_i\rangle|^2 \\
	&=\argmax{\vecx\in\List'} \sum_{i=1}^{\ell} \Re\{\langle\vecyd_i,\hat{h}_i\vecxd_i\rangle\}+\frac{1}{2n_p}|\langle\vecyd_i,\vecxd_i\rangle|^2.\label{eq:GLRT_inList}
\end{align}

This metric lends itself to an alternative interpretation. Suppose that the distribution of the channel coefficient for the $i$th coherence block is a complex Gaussian distribution with mean $\hat{h}_i$ given in \eqref{eq:channel_estimate} and variance $\frac{2\sigma^2}{\norm{\vecxp_i}^2}$, i.e., $H_i\sim\mathcal{CN}\left(\hat{h}_i,\frac{2\sigma^2}{\norm{\vecxp_i}^2}\right)$. Then, similar to \eqref{eq:nc_ML_def}, we obtain
\begin{align}
	\hat{\vecx} &= \argmax{\vecx\in\List'} \prod_{i=1}^{\ell} \mathbb{E}[p_{\randvecyd|\randvecxd,H_i}(\vecyd_i|\vecxd_i,H_i)] \\
	&= \argmax{\vecx\in\List'} \sum_{i=1}^{\ell} \frac{\norm{\vecxd_i}^2+|\langle\vecyd_i,\vecxd_i\rangle|^2+2\norm{\vecxp_i}^2\Re\{\langle\vecyd_i,\hat{h}_i\vecxd_i\rangle\}}{\norm{\vecxp_i}^2+\norm{\vecxd_i}^2} \nonumber\\
	&\qquad-|\hat{h}_i|^2\norm{\vecxd_i}^2+\sigma^2\ln\left(\frac{\norm{\vecxp_i}^2}{\norm{\vecxp_i}^2+\norm{\vecxd_i}^2}\right)
\end{align}
where $\code'$ is the modified channel code obtained by re-inserting the pilot symbols to each codeword. By assuming \ac{QPSK} (which implies $\norm{\vecxp_i}^2=n_p$), we recover \eqref{eq:GLRT_inList}. Note that the decoding metric has two contributions: A first term that resembles a coherent metric based on the estimate $\hat{\vech}$, and a second term that is related to the non-coherent correlation. The second term is weighted by the inverse of the number of pilots; hence it becomes negligible when $n_p$ is large (i.e., when the channel estimate is reliable).
	
\section{Finite-blocklength Bounds}
\label{sec:bounds}

We review the converse and achievability bounds on the average error probability based on finite-blocklength information theory that will be used to benchmark the coding schemes introduced in the previous section.
The converse bound is based on the metaconverse theorem in \cite[Thm. 28]{polyanskiy10-05a} and the achievability bounds are based on the \ac{RCUs} \cite[Thm. 1]{martinez11-02a}.

Let $q:\mathbb{C}^{\nc} \times \mathbb{C}^{\nc} \rightarrow \mathbb{R}^+$ be an arbitrary block-wise decoding metric and let $(\bar{\randvecx}_i, \randvecx_i, \randvecy_i)\sim  p_{\randvecx}(\bar{\vecx}_i) p_{\randvecx}(\vecx_i) p_{\randvecy | \randvecx}(\vecy_i  |\vecx_i)$, $i=1,\dots, \ell$, be independent across coherence blocks.
The generalized information density is defined as
\begin{IEEEeqnarray}{rCl} 
	\imath_s(\vecx_i, \vecy_i) \triangleq \ln\frac{q(\vecx_i, \vecy_i)^s}{\mathbb{E}[q(\bar{\randvecx}_i, \vecy_i)^s]}
\end{IEEEeqnarray}
where $s\geq 0$. The RCUs achievability bound states that, for a given rate $R$, the average error probability is upper-bounded as
\begin{IEEEeqnarray}{rCl} 
	\epsilon &\leq& 
	\epsilon_{\sss{rcus}} \\ &\triangleq & \inf_{s\geq 0}\mathbb{E}\left[ e^{- \lrho{ \sum_{i=1}^\ell\imath_s(\randvecx_i,\randvecy_i) - \ln(2^{R\nc\ell}-1) }^+} \right]. \label{eq:rcus}
\end{IEEEeqnarray}
We evaluate the bound in~\eqref{eq:rcus} for the following combinations of input distributions and decoding metrics:
\begin{itemize}
\item[i)]  Input symbols uniformly distributed on a shell in $\mathbb{C}^{n_c}$, and \ac{ML}  decoding, i.e., $q(\vecx_i, \vecy_i) = p_{\randvecy |\randvecx}(\vecy_i |\vecx_i)$;
\item[ii)] a pilot-assisted scheme as in Section~\ref{subsec:complexity} with the $n_c-n_p$ data symbols uniformly distributed on a shell in $\mathbb{C}^{n_c-n_p}$ and ML decoding, i.e., $q(\vecx_i, \vecy_i) = p_{\randvecy^{\sss{d}}  |\randvecx^{\sss{d}}, \hat{H}}(\vecyd_i  |\vecxd_i, \hat{h}_i)$;
\item[iii)] input distribution as in ii), and scaled nearest neighbor decoding, i.e., $q(\vecx_i, \vecy_i) = \exp(-\vecnorm{\vecyd_i-\hat{h}_i\vecxd_i}^2)$.
\end{itemize}
See \cite[Sec. III.A-III.D]{johan:unpublished} for additional details on how to evaluate~\eqref{eq:rcus} for each of these cases.

Next, we state the converse bound, which is based on the metaconverse theorem,
For a given average error probability $\epsilon$, the maximum code rate is upper-bounded as
\begin{IEEEeqnarray}{rCl} 
R &\leq & R_{\sss{mc}}\lro{\epsilon} \\
& \triangleq & \inf_{\lambda\geq 0} \frac{1}{\ell\nc}\lro{\lambda - \ln\lrho{\prob{\sum_{i=1}^\ell \imath_1(\randvecx_i,\randvecy_i) \leq \lambda} - \epsilon}^+ }.\IEEEeqnarraynumspace\label{eq:metaconverse}
\end{IEEEeqnarray}
For a given rate $R$, a lower bound on $\epsilon$, denoted as $\epsilon_{\sss{mc}}$, can be obtained from~\eqref{eq:metaconverse} by finding the $\epsilon_{\sss{mc}}$ for which $ R_{\sss{mc}}\lro{\epsilon_{\sss{mc}}} = R$.
For more details on this converse bound, the reader is referred to \cite[Sec. III.E]{johan:unpublished}.

	\section{Numerical Results}\label{sec:results}

We present next an example of the performance achieved by the decoder proposed in Section \ref{sec:opt_in_list}. The results are obtained by Monte Carlo simulations and are provided in terms of \ac{BLER} vs. \ac{SNR} with the \ac{SNR} espressed as $E_s/N_0$, where $E_s$ is the expected energy per symbol and $N_0$ the single-sided noise power spectral density. The results are compared with the bounds of Section \ref{sec:bounds}. We consider a Rayleigh block-fading channel with $4$ diversity branches. Each branch consists of $13$ channel uses, which results in $52$ channel uses per message.
For the simulations, we considered the case where $k=32$ information bits are transmitted within each codeword, yielding a rate $R=32/52 \approx 0.62$ bits per channel use.
The symbols are taken from a \ac{QPSK} constellation. A $\left(96, 32\right)$ quasi-cyclic code is used to transmit and a suitable number of codeword bits is punctured to accommodate the pilot symbols within the $52$ channel uses. The code is obtained by tail-biting termination of a rate$-1/3$ non-systematic convolutional code with a memory $17$ and generators $[552137, 614671, 772233]$ \cite[Table 10.14]{johannesson_book}. The  minimum distance of the quasi-cyclic code is upper-bounded by the free distance of the underlying convolutional code, which is $32$. After encoding, a pseudo-random interleaver is applied to the codeword bits. Then puncturing adapts the blocklength to the number of channel uses available after pilot insertion.
The \ac{OSD} order is set to $3$, which provides a reasonable trade-off between performance and decoding complexity. With this choice, \ac{OSD} builds a list $\mathcal{L}$ of $\sum_{i=0}^{3}{k\choose i} = 5489$ candidate codewords.

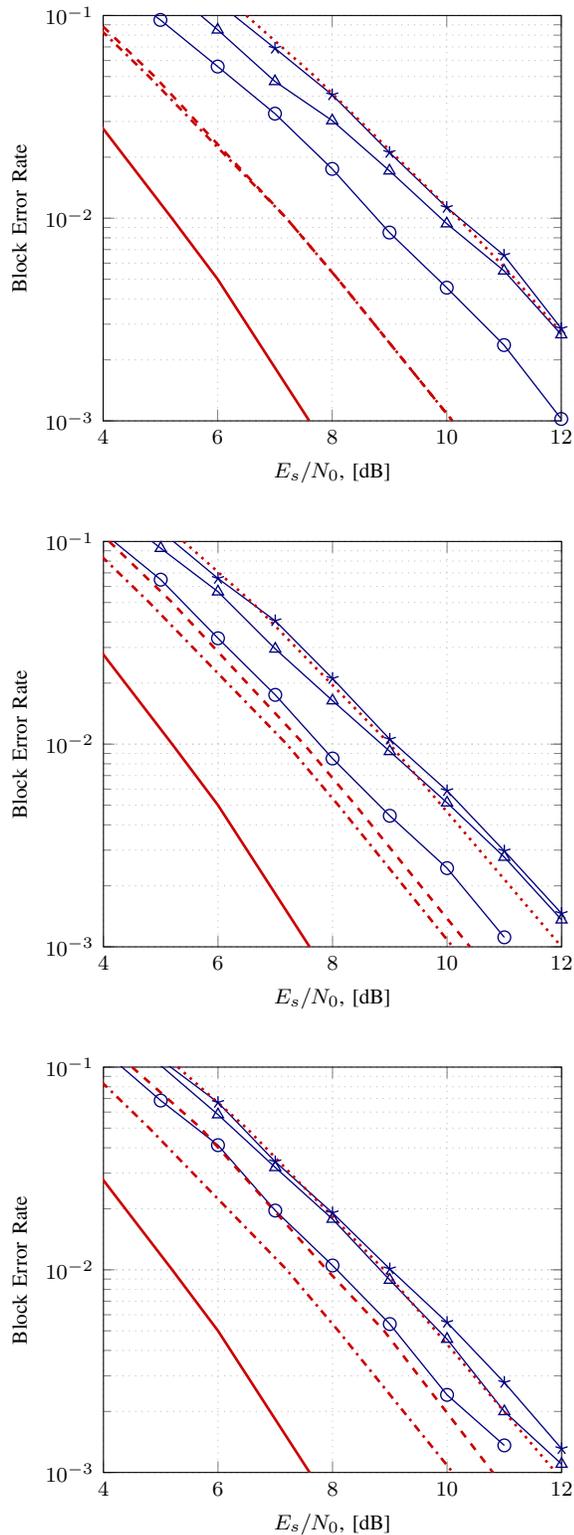
\begin{figure}    
	\centering
	\begin{subfigure}{0.875\columnwidth}
		\begin{tikzpicture}[font=\footnotesize]
\begin{semilogyaxis}[
	 mark options={solid,scale=1.2},
     width=0.99*\columnwidth,
     height=0.9*\columnwidth,
     grid=both,
     grid style={dotted,gray!50},
    ylabel near ticks,
    xlabel near ticks,
	xmin=4,
	xmax=12,
	ymin=1e-3,
	ymax=1e-1,
	xlabel={$E_s/N_0$, [dB]},
	ylabel={Block Error Rate}
	]

\addplot[color=blue!50!black,line width = 0.5pt,solid,mark=star] table[x=snr,y=P] {figures/96_32np1.txt}; \label{leg:pragmaticnp1}
\addplot[color=blue!50!black,line width = 0.5pt,solid,mark=o] table[x=snr,y=P] {figures/96_32_opt_np1.txt}; \label{leg:optnp1}
\addplot[color=blue!50!black,line width = 0.5pt,solid,mark=triangle] table[x=snr,y=P] {figures/96_32np1EM_v2.txt}; \label{leg:pragmaticnp1EM}

\addplot[color=red!80!black,line width = 1pt,dotted] table[x=snr,y=P] {figures/52_32_NN_np1_ach.txt}; \label{leg:NNach52_32_np1}

\addplot[color=red!80!black,line width = 1pt,dashed] table[x=snr,y=P] {figures/52_32_ML_np1_ach.txt}; \label{leg:ach52_32_np1}

\addplot[color=red!80!black,line width = 1pt,dashdotted] table[x=snr,y=P] {figures/52_32_ML_np0_ach.txt}; \label{leg:ach52_32_np0}

\addplot[color=red!80!black,line width = 1pt,solid] table[x=snr,y=P] {figures/52_32_ML_conv.txt}; \label{leg:conv52_32}

\end{semilogyaxis}
\end{tikzpicture}
	\end{subfigure}
	\begin{subfigure}{0.875\columnwidth}
		\begin{tikzpicture}[font=\footnotesize]
\begin{semilogyaxis}[
	 mark options={solid,scale=1.2},
     width=0.99*\columnwidth,
     height=0.9*\columnwidth,
     grid=both,
     grid style={dotted,gray!50},
    ylabel near ticks,
    xlabel near ticks,
	xmin=4,
	xmax=12,
	ymin=1e-3,
	ymax=1e-1,
	xlabel={$E_s/N_0$, [dB]},
	ylabel={Block Error Rate}
	]

\addplot[color=blue!50!black,line width = 0.5pt,solid,mark=star] table[x=snr,y=P] {figures/96_32np2.txt}; \label{leg:pragmaticnp2} 
\addplot[color=blue!50!black,line width = 0.5pt,solid,mark=o] table[x=snr,y=P] {figures/96_32_opt_np2.txt}; \label{leg:optnp2}
\addplot[color=blue!50!black,line width = 0.5pt,solid,mark=triangle] table[x=snr,y=P] {figures/96_32np2EM.txt}; \label{leg:pragmaticnp2EM}
\addplot[color=red!80!black,line width = 1pt,dotted] table[x=snr,y=P] {figures/52_32_NN_np2_ach.txt}; \label{leg:NNach52_32_np2}

\addplot[color=red!80!black,line width = 1pt,dashed] table[x=snr,y=P] {figures/52_32_ML_np2_ach.txt}; \label{leg:ach52_32_np2}

\addplot[color=red!80!black,line width = 1pt,dashdotted] table[x=snr,y=P] {figures/52_32_ML_np0_ach.txt}; \label{leg:ach52_32_np0}

\addplot[color=red!80!black,line width = 1pt,solid] table[x=snr,y=P] {figures/52_32_ML_conv.txt}; \label{leg:conv52_32}

\end{semilogyaxis}
\end{tikzpicture}
	\end{subfigure}
	\begin{subfigure}{0.875\columnwidth}
		\begin{tikzpicture}[font=\footnotesize]
\begin{semilogyaxis}[
	 mark options={solid,scale=1.2},
     width=0.99*\columnwidth,
     height=0.9*\columnwidth,
     grid=both,
     grid style={dotted,gray!50},
    ylabel near ticks,
    xlabel near ticks,
	xmin=4,
	xmax=12,
	ymin=1e-3,
	ymax=1e-1,
	xlabel={$E_s/N_0$, [dB]},
	ylabel={Block Error Rate}
	]

\addplot[color=blue!50!black,line width = 0.5pt,solid,mark=star] table[x=snr,y=P] {figures/96_32np3.txt}; \label{leg:pragmaticnp3}
\addplot[color=blue!50!black,line width = 0.5pt,solid,mark=o] table[x=snr,y=P] {figures/96_32_opt_np3.txt}; \label{leg:optnp3}
\addplot[color=blue!50!black,line width = 0.5pt,solid,mark=triangle] table[x=snr,y=P] {figures/96_32np3EM.txt}; \label{leg:pragmaticnp3EM}

\addplot[color=red!80!black,line width = 1pt,dotted] table[x=snr,y=P] {figures/52_32_NN_np3_ach.txt}; \label{leg:NNach52_32_np3}

\addplot[color=red!80!black,line width = 1pt,dashed] table[x=snr,y=P] {figures/52_32_ML_np3_ach.txt}; \label{leg:ach52_32_np3}

\addplot[color=red!80!black,line width = 1pt,dashdotted] table[x=snr,y=P] {figures/52_32_ML_np0_ach.txt}; \label{leg:ach52_32_np0}

\addplot[color=red!80!black,line width = 1pt,solid] table[x=snr,y=P] {figures/52_32_ML_conv.txt}; \label{leg:conv52_32}

\end{semilogyaxis}
\end{tikzpicture}
	\end{subfigure}
	\caption{BLER vs. SNR for the proposed scheme \eqref{leg:optnp1} with $n_p=1$ (top), $n_p=2$ (middle) and $n_p=3$ (bottom). Finite length performance bounds given by the converse bound of \eqref{eq:metaconverse} \eqref{leg:conv52_32}, the achievability of \eqref{eq:rcus} for a non-coherent setup with \ac{ML} decoding \eqref{leg:ach52_32_np0}, for \ac{PAT} under \ac{ML} decoding \eqref{leg:ach52_32_np1} and for \ac{PAT} under scaled nearest neighbor decoding \eqref{leg:NNach52_32_np1}. The performance of the pragmatic \ac{PAT} \ac{OSD} scheme of Section \ref{sec:PATOSD} \eqref{leg:pragmaticnp1} and the performance of the \ac{EM}-based approach \eqref{leg:pragmaticnp1EM} is provided as a reference.
	}\label{fig:combined}
\end{figure}

In Fig. \ref{fig:combined}, we compare the performance of the proposed decoder to the performance of the  two baseline decoders described in Section \ref{sec:pragmatic}  for different numbers of pilot symbols ($n_p \in \{1,2,3\}$) per coherence block. For the iterative \ac{EM}-based \ac{OSD}, we set the number of iteration to $m=1$. For the tested cases, the gain achieved by the proposed decoder is up to $1.2$ dB as compared to the simple pilot-aided \ac{OSD} of Section \ref{sec:PATOSD}. The performance of the iterative \ac{EM}-based \ac{OSD} is only marginally better than the one obtained by the simple pilot-aided \ac{OSD}. Remarkably, the proposed approach performs close to the \ac{RCUs} for \ac{PAT} and ML decoding except for $n_p = 1$. In the simulated setting, the proposed approach provides the best performance with $n_p=2$, with a slight degradation visible when $n_p=3$.
	\section{Conclusions}\label{sec:conc}

We proposed a novel decoding method over fading channels with no \ac{CSI} at the transmitter/receiver, which leverages an initial (rough) pilot-assisted channel estimate to construct a list, and then performs the final decision by applying a non-coherent decoding metric to the list elements. The approach can be applied to codes that admit list decoding. We demonstrated its application to \ac{OSD}, and showed that, in the short blocklength regime, it is possible to operate close to tight random coding achievability bounds.
	
	\section*{Acknowledgement}
	The authors would like to thank Gerhard Kramer for the helpful comments
	that improved the presentation of this paper.


\begin{thebibliography}{10}
\providecommand{\url}[1]{#1}
\csname url@samestyle\endcsname
\providecommand{\newblock}{\relax}
\providecommand{\bibinfo}[2]{#2}
\providecommand{\BIBentrySTDinterwordspacing}{\spaceskip=0pt\relax}
\providecommand{\BIBentryALTinterwordstretchfactor}{4}
\providecommand{\BIBentryALTinterwordspacing}{\spaceskip=\fontdimen2\font plus
\BIBentryALTinterwordstretchfactor\fontdimen3\font minus
  \fontdimen4\font\relax}
\providecommand{\BIBforeignlanguage}[2]{{%
\expandafter\ifx\csname l@#1\endcsname\relax
\typeout{** WARNING: IEEEtran.bst: No hyphenation pattern has been}%
\typeout{** loaded for the language `#1'. Using the pattern for}%
\typeout{** the default language instead.}%
\else
\language=\csname l@#1\endcsname
\fi
#2}}
\providecommand{\BIBdecl}{\relax}
\BIBdecl

\bibitem{Durisi16:Short}
G.~Durisi, T.~Koch, and P.~Popovski, ``Towards massive, ultra-reliable, and
  low-latency wireless communications with short packets,'' \emph{Proc.
  {IEEE}}, vol. 104, no.~9, pp. 1711--1726, Sep. 2016.

\bibitem{Dolinar98:BOUNDS}
S.~Dolinar, D.~Divsalar, and F.~Pollara, ``Code performance as a function of
  block size,'' Jet Propulsion Laboratory, Pasadena, CA, USA, TMO progress
  report 42-133, May 1998.

\bibitem{Fossorier04:BOUNDS}
A.~Valembois and M.~Fossorier, ``{Sphere-Packing Bounds Revisited for Moderate
  Block Lengths},'' \emph{{IEEE} Trans. Inf. Theory}, vol.~50, no.~12, pp. 2998
  -- 3014, Dec. 2004.

\bibitem{SasonShamai06:BOUNDS}
I.~Sason and S.~Shamai, \emph{Performance Analysis of Linear Codes under
  Maximum-Likelihood Decoding: A Tutorial}.\hskip 1em plus 0.5em minus
  0.4em\relax Delft, The Netherlands: Now Publisher Inc., Jul. 2006, vol.~3,
  no. 1--2.

\bibitem{polyanskiy10-05a}
Y.~Polyanskiy, H.~V. Poor, and S.~Verd\'u, ``Channel coding rate in the finite
  blocklength regime,'' \emph{{IEEE} Trans. Inf. Theory}, vol.~56, no.~5, pp.
  2307--2359, May 2010.

\bibitem{Poulliat08:BinImag}
C.~Poulliat, M.~Fossorier, and D.~Declercq, ``{Design of regular $(2,
  d_c)$-{LDPC} codes over {GF}(q) using their binary images},'' \emph{{IEEE}
  Trans. Commun.}, vol.~56, no.~10, pp. 1626--1635, 2008.

\bibitem{Liva13:ShortTC}
G.~Liva, E.~Paolini, B.~Matuz, S.~Scalise, and M.~Chiani, ``Short turbo codes
  over high order fields,'' \emph{{IEEE} Trans. Commun.}, vol.~61, no.~6, pp.
  2201--2211, June 2013.

\bibitem{Dolecek14:NBLDPCTIT}
L.~Dolecek, D.~Divsalar, Y.~Sun, and B.~Amiri, ``Non-binary protograph-based
  {LDPC} codes: Enumerators, analysis, and designs,'' \emph{{IEEE} Trans. Inf.
  Theory}, vol.~60, no.~7, pp. 3913--3941, July 2014.

\bibitem{FosLin95}
M.~P.~C. Fossorier and S.~Lin, ``Soft-decision decoding of linear block codes
  based on ordered statistics,'' \emph{Trans. on Inf. Theory}, vol.~41, no.~5,
  pp. 1379--1396, Sep. 1995.

\bibitem{Tal15:ListPolar}
I.~Tal and A.~Vardy, ``List decoding of polar codes,'' \emph{{IEEE} Trans. Inf.
  Theory}, vol.~61, no.~5, pp. 2213--2226, May 2015.

\bibitem{Coskun18:Survey}
\BIBentryALTinterwordspacing
M.~C. Co\c{s}kun, G.~Durisi, T.~Jerkovits, G.~Liva, W.~Ryan, B.~Stein, and
  F.~Steiner, ``Efficient error-correcting codes in the short blocklength
  regime,'' \emph{CoRR}, vol. abs/1706.05238, 2018. [Online]. Available:
  \url{https://arxiv.org/abs/1812.08562}
\BIBentrySTDinterwordspacing

\bibitem{Tong04:PAT}
L.~Tong, B.~M. Sadler, and M.~Dong, ``Pilot-assisted wireless transmissions:
  General model, design criteria, and signal processing,'' \emph{{IEEE} Signal
  Process. Mag.}, vol.~21, no.~6, pp. 12--25, Nov. 2004.

\bibitem{Liva17:Mismatched}
G.~Liva, G.~Durisi, M.~Chiani, S.~S. Ullah, and S.~C. Liew, ``{Short codes with
  mismatched channel state information: A case study},'' in \emph{IEEE Int.
  Workshop on Signal Process. Adv. in Wireless Commun. (SPAWC)}, Sapporo,
  Japan, Jul 2017, pp. 1--5.

\bibitem{Yang14:QSFC}
W.~Yang, G.~Durisi, T.~Koch, and Y.~Polyanskiy, ``Quasi-static multiple-antenna
  fading channels at finite blocklength,'' \emph{{IEEE} Trans. Commun.},
  vol.~60, no.~7, pp. 4232--4265, July 2014.

\bibitem{Durisi16:MRFC}
G.~Durisi, T.~Koch, J.~\"{O}stman, Y.~Polyanskiy, and W.~Yang, ``Short-packet
  communications over multiple-antenna {R}ayleigh-fading channels,''
  \emph{{IEEE} Trans. Commun.}, vol.~64, no.~2, pp. 618--629, Feb 2016.

\bibitem{östman17-02a}
J.~\"{O}stman, G.~Durisi, E.~G. Str\"om, J.~Li, H.~Sahlin, and G.~Liva,
  ``Low-latency ultra-reliable {5G} communications: Finite block-length bounds
  and coding schemes,'' in \emph{Int. {ITG} Conf. Sys. Commun. Coding (SCC)},
  Hamburg, Germany, Feb. 2017.

\bibitem{johan:unpublished}
\BIBentryALTinterwordspacing
J.~\"Ostman, G.~Durisi, E.~G. Str\"om, M.~C. Co\c{s}kun, and G.~Liva, ``Short
  packets over block-memoryless fading channels: Pilot-assisted or noncoherent
  transmission?'' \emph{{IEEE} Trans. Commun.}, 2018, to appear. [Online].
  Available: \url{http://arxiv.org/pdf/1712.06387.pdf}
\BIBentrySTDinterwordspacing

\bibitem{Warrier02}
D.~Warrier and U.~Madhow, ``Spectrally efficient noncoherent communication,''
  \emph{{IEEE} Trans. Inf. Theory}, vol.~48, no.~3, pp. 651--668, Mar. 2002.

\bibitem{henkbook}
H.~Wymeersch, \emph{Iterative Receiver Design}.\hskip 1em plus 0.5em minus
  0.4em\relax Cambridge: Cambridge University Press, 2007.

\bibitem{herzet2007code}
C.~Herzet, N.~Noels, V.~Lottici, H.~Wymeersch, M.~Luise, M.~Moeneclaey, and
  L.~Vandendorpe, ``Code-aided turbo synchronization,'' \emph{Proc. of the
  IEEE}, vol.~95, no.~6, pp. 1255--1271, 2007.

\bibitem{KhaBout:EM-APP06}
M.~Khalighi and J.~J. Boutros, ``Semi-blind channel estimation using the {EM}
  algorithm in iterative {MIMO} {APP} detectors,'' \emph{{IEEE} Trans. Wireless
  Commun.}, vol.~5, no.~11, pp. 3165--3173, November 2006.

\bibitem{Dempster77maximumlikelihood}
A.~P. Dempster, N.~M. Laird, and D.~B. Rubin, ``Maximum likelihood from
  incomplete data via the {EM} algorithm,'' \emph{Journal of the Royal
  statistical society, series B}, vol.~39, no.~1, pp. 1--38, 1977.

\bibitem{martinez11-02a}
A.~Martinez and A.~{Guill{\'e}n i F{\`a}bregas}, ``Saddlepoint approximation of
  random--coding bounds,'' in \emph{Proc. Inf. Theory Applicat. Workshop
  (ITA)}, San Diego, CA, U.S.A., Feb. 2011.

\bibitem{johannesson_book}
R.~Johannesson and K.~S. Zigangirov, \emph{Fundamentals of Convolutional
  Coding}, 2nd~ed.\hskip 1em plus 0.5em minus 0.4em\relax Piscataway, NJ, USA:
  Wiley-IEEE Press, 2015.

\end{thebibliography}

\end{document}